\documentclass[twocolumn,showpacs,preprintnumbers,amsmath,amssymb]{revtex4}

\usepackage{graphicx}
\usepackage{dcolumn}
\usepackage{bm}

\begin{document}

\title{Vacuum Rabi oscillation of an atom without rotating-wave approximation}

\author{Fa-Qiang Wang}
\email{fqwang@scnu.edu.cn}
\author{Wei-Ci Liu}
\author{Rui-Sheng Liang}%
\affiliation{Lab of Photonic Information Technology, School of Information and Photoelectronic Science and Engineering, South China Normal University, Guangzhou 510006, China}%

\date{\today}

\begin{abstract}
We have investigated vacuum Rabi oscillation
 of an atom coupled with single-mode cavity field exactly, and
 compared the results with that of J-C model. The results show that, for resonant case, there is
  damping Rabi oscillation for an atom, even in strong coupling regime. For small
 detuning and weak coupling case, the probability for the atom in excited
 state oscillates against time with different frequency and amplitude
 from that of J-C model. It exhibits the counter-rotating wave interaction could significantly effect
 the dynamic behavior of the atom, even under the condition in which the RWA is considered to be justified.
 On the other hand, the results also reveal that there is Rabi oscillation for initially unexcited atom,
 which is contrary to that of J-C model.

\end{abstract}

\pacs{42.50.Md, 31.70.Hq, 42.50.Lc}
\maketitle

\par A two-level atom interacting with a single cavity mode,
described by the Jaynes-Cummings(J-C) model, which neglects counter
rotating terms, is widely used in quantum optics and has the
potential to constitute the basic building block of quantum
computers\cite{Lou,pur,kli,rai}. Making rotating-wave
approximation(RWA) in Hamiltonian strongly simplifies the
mathematical treatment of the problem and usually give exact
solution of the approximate Hamiltonian. In spite of the simplicity
of the J-C model, the dynamics have turned out to be various and
complex, describing many physical phenomena, such as Rabi
oscillations, collapse-revivals, squeezing, atom-field
entanglement\cite{Ger}.

\par Generally, the RWA, which neglecting counter rotating, is justified
for small detuning and small  ratio of the atom-field coupling
divided by the atomic transition frequency. In atom-field  cavity
systems, this ratio is typically of the order  $10^{-7}\sim10^{-6}$.
Recently, cavity systems with very strong couplings have been
discussed\cite{mei}. The ratio might become order of magnitudes
larger in solid state systems and the counter-rotating wave terms
must be considered\cite{iri}. In this paper, we investigate the
influence of counter-rotating wave terms on the decay behavior of an
atoms coupled with one-mode cavity, without rotating-wave
approximation.

\par Now we restrict our attention to a two-level
atom coupled with a perfect one-mode cavity field, of which the
Hamiltonian is
\begin{equation}\label{e1}
    H=H_{a}+H_{f}+H_{af}
\end{equation}
where
\begin{eqnarray}
  H_{a} &=& \omega_{0}\frac{\sigma_{z}}{2} \\
  H_{f} &=& \omega a^{\dagger}a \\
  H_{af} &=& g(\sigma_{+}+\sigma_{-})(a^{\dagger}+a)
\end{eqnarray}
where $\omega_{0}$ is the atomic transition frequency between the
ground state $|0\rangle$ and excited state $|1\rangle$.
$\sigma_{z}=|1\rangle\langle1|-|0\rangle\langle0|$,
$\sigma_{+}=|1\rangle\langle0|$ and $\sigma_{-}=|0\rangle\langle1|$
are pseudo-spin operators of atom. $a^{\dagger}$ and $a$ are
creation and annihilation operators of the cavity field mode
corresponding frequency $\omega$. And $g$ is the coupling constant
between the transition $|1\rangle-|0\rangle$ and the field mode.

\par If the cavity field is initially in vacuum state
, the non-perturbative reduced master equation of the atom could be
derived by path integrals\cite{Ish}
\begin{eqnarray}\label{e7}
  \frac{\partial}{\partial
    t}\rho_{a} &=& \left[\varepsilon_{0}J_{0}+ \varepsilon_{+}J_{+}+\varepsilon_{-}J_{-})\right]\rho_{a}-g^{2}\left[\alpha^{R}+f\right]\rho_{a}\nonumber\\
   & & +\left[\nu_{0}K_{0}+\nu_{+}K_{+}+\nu_{-}K_{-}\right]\rho_{a}
\end{eqnarray}
Where $\varepsilon_{0}=-2i\left(\omega_{0}-g^{2}
\alpha^{I}+g^{2}f^{I}\right) $,
$\varepsilon_{+}=g^{2}\left(\alpha+f^{*}\right)$,
$\varepsilon_{-}=g^{2}\left(\alpha^{*}+f\right)$, $\nu_{0}=2g^{2}
\left(\alpha^{R}-f^{R}\right)$, $ \nu_{+}=2g^{2}\alpha^{R}$,
$\nu_{-}=2g^{2} f^{R}$. $J_{0}$, $J_{+}$, $J_{-}$, $K_{0}$, $K_{+}$
and $K_{-}$ are superoperators defined as
\begin{eqnarray*}
  J_{0}\rho_{a} &\equiv& \left[\frac{\sigma_{z}}{4},\rho_{a}\right] \\
  J_{+}\rho_{a} &\equiv& \sigma_{+}\rho_{a}\sigma_{+} \\
  J_{-}\rho_{a} &\equiv&  \sigma_{-}\rho_{a}\sigma_{-} \\
  K_{0}\rho_{a} &\equiv& (\sigma_{+}\sigma_{-}\rho_{a}+\rho_{a}\sigma_{+}\sigma_{-}-\rho_{a})/2 \\
  K_{+}\rho_{a} &\equiv& \sigma_{+}\rho_{a}\sigma_{-} \\
  K_{-}\rho_{a} &\equiv& \sigma_{-}\rho_{a}\sigma_{+}
\end{eqnarray*}
and
\begin{eqnarray}\label{e71}
  \alpha &=& \frac{1-exp(-i\Delta t)}{i\Delta} \\
  f &=& \frac{exp(i\delta t)-1}{i\delta}
\end{eqnarray}
where $\Delta=\omega+\omega_{0}$, $\delta=\omega_{0}-\omega$.
$\alpha^{R}$, $\alpha^{I}$, $\alpha^{*}$ and $f^{R}$, $f^{I}$,
$f^{*}$ are real part, image part and conjugate of $\alpha$ and of
$f$, respectively. $\alpha$ comes from the counter-rotating wave
interaction and $f$ comes from the rotating-wave interaction.

\par Using algebraic approach, the formal solution of Eq.(\ref{e7}) is
obtained \cite{pur,zhang}
\begin{eqnarray}\label{e8}
  \rho_{a}(t) &=& e^{-\Gamma_{k}}\hat{T}e^{\int_{0}^{t}dt(\varepsilon_{0}J_{0}+ \varepsilon_{+}J_{+}+\varepsilon_{-}J_{-})}\nonumber\\
    & & \times \hat{T}e^{\int_{0}^{t}dt(\nu_{0}K_{0}+\nu_{+}K_{+}+\nu_{-}K_{-})}\rho_{a}(0)
\end{eqnarray}
where  $\Gamma_{k}=g^{2}(\tilde{\alpha}^{R}+F^{R})$ and
\begin{eqnarray}\label{e9}
  \tilde{\alpha} &=& \int_{0}^{t}\alpha dt =\frac{1-exp(-i\Delta t)-i\Delta t}{\Delta^{2}} \equiv\tilde{\alpha}^{R}+i\tilde{\alpha}^{I} \nonumber\\
  F &=& \int_{0}^{t}f dt =\frac{1+i\delta t-exp(i\delta t)}{\delta^{2}} \equiv F^{R}+iF^{I}
\end{eqnarray}
where $\tilde{\alpha}^{R}$, $\tilde{\alpha}^{I}$,
$\tilde{\alpha}^{*}$ and $F^{R}$, $F^{I}$, $F^{*}$ are real part,
image part and conjugate of $\tilde{\alpha}$ and of $F$,
respectively.

\par Using the decomposition of SU(2) operator, the time-ordered exponential operators could be disentangled\cite{pur}.
The exact solution of master equation Eq.(\ref{e7}) is obtained
\begin{equation}\label{e10}
     \rho_{a}(t) = e^{-\Gamma_{k}}\tilde{\rho}(t)
\end{equation}
\begin{equation}\label{e11}
    \tilde{\rho}(t)=\left(\begin{array}{cc}
                           l\rho^{11}_{a}(0)+ m\rho^{00}_{a}(0) & x\rho^{10}_{a}(0)+ y\rho^{01}_{a}(0) \\
                           q\rho^{01}_{a}(0)+ r\rho^{10}_{a}(0) &
                           n\rho^{00}_{a}(0)+ p\rho^{11}_{a}(0)
                         \end{array}
                         \right)
\end{equation}
\begin{eqnarray}
  l &=& e^{k_{0}/2}+e^{-k_{0}/2}k_{+}k_{-},\ m = e^{-k_{0}/2}k_{+} \\
  n &=& e^{-k_{0}/2},\ p= e^{-k_{0}/2}k_{-}\\
  q &=& e^{-j_{0}/2},\ r= e^{-j_{0}/2}j_{-} \\
  x &=& e^{j_{0}/2}+e^{-j_{0}/2}j_{+}j_{-},\ y=e^{-j_{0}/2}j_{+}
\end{eqnarray}
where $j_{+}$, $j_{0}$, $j_{-}$ and $k_{+}$, $k_{0}$, $k_{-}$
satisfy the following differential equation\cite{pur}
\begin{eqnarray}
  \dot{X}_{+} &=& \mu_{+}-\mu_{-}X_{+}^{2}+\mu_{0}X_{+} \\
  \dot{X}_{0} &=& \mu_{0}-2\mu_{-}X_{+}\\
  \dot{X}_{-} &=& \mu_{-}exp(X_{0})
\end{eqnarray}
$\mu=\varepsilon$ for $X=j$ and $\mu=\nu$ for $X=k$. Generally, the
Riccati equation could not be solved analytically. Next, we will
investigate it numerically.

\par When the atom is initially in excited state $\rho^{11}_{a}(0)=1$, the probability $P_{e}$ for
the atom in excited state at time $t$ is $P_{e}=l$. The counterpart
result for Jaynes-Cummings model is
\begin{equation}\label{e14}
    P^{JC}_{e}=1-[\frac{2gsin(\Omega t/2)}{\Omega}]^{2}
\end{equation}
where $\Omega=\sqrt{\delta^{2}+4g^{2}}$

\par When the atom is initially in ground state $\rho^{00}_{a}(0)=1$, the probability $P_{e}$ for
the atom in excited state at time $t$ is $P_{e}=m$. The counterpart
result for Jaynes-Cummings model is $ P^{JC}_{e}=0$.

 \par First, we investigate the case for initially excited atom. And
 we refer to our exact solution as exact model in the following discussion.

\par
\begin{figure}
  \includegraphics{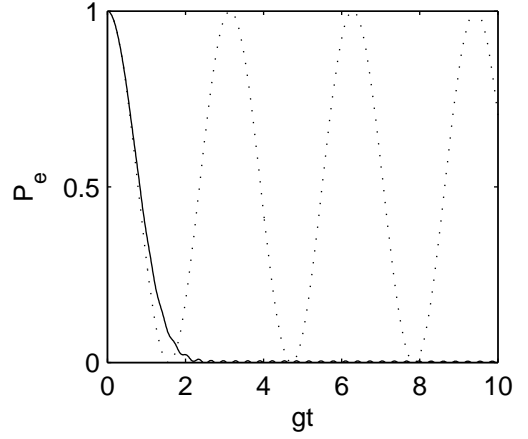}\\
  \caption{$P_{e}$ as a function of $gt$ for initially excited atom with $\omega_{0}=10g$ and $\delta=0$. dotted line for J-C model, solid line for exact model.}\label{F1}
\end{figure}

 (A) For resonant and weak coupling case, Fig.\ref{F1} reveals that
 the probability $P_{e}$, for J-C model, oscillates against time, while that
 periodically decays and revives with damping amplitude for exact model.
 Eq.(\ref{e10}) and (\ref{e14}) show that there is an attenuation factor $exp(-g^{2}t^{2}/2)$ for exact model when
$\omega_{0}t\gg1$, while $P^{JC}_{e}=1-sin^{2}(gt)$ for J-C model.

\begin{figure}
  \includegraphics{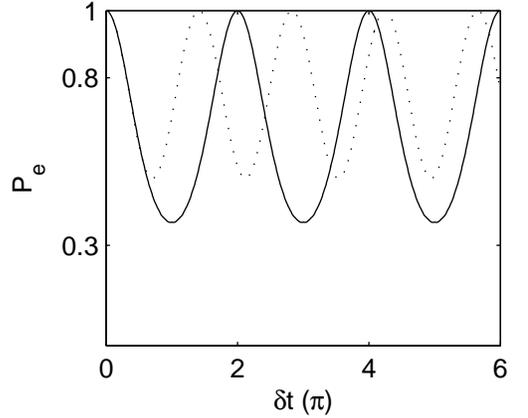}\\
  \caption{$P_{e}$ as a function of $\delta t$ for initially excited atom with $\omega_{0}=20g$ and $\delta=0.1\omega_{0}$. dotted line for J-C model, solid line for exact model.}\label{F2}
\end{figure}

\par (B)For small detuning and weak coupling case, Fig(\ref{F2})
shows that $P_{e}$ oscillates against time for the two model except
for different amplitude and different oscillating frequency. For
exact model, the result shows that the oscillating frequency is
$\delta$ for exact model, while that is $\Omega$ for J-C model.
\par
\begin{figure}
  \includegraphics{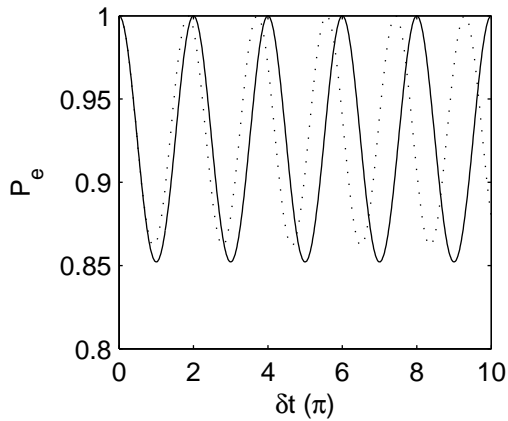}\\
  \caption{$P_{e}$ as a function of $\delta t$ for initially excited atom with $\omega_{0}=50g$ and $\delta=0.1\omega_{0}$. dotted line for J-C model, solid line for exact model.}\label{F3}
\end{figure}

 From Fig.(\ref{F3}), we could find, with the decreasing of the ratio of
coupling strength to the atom transmission frequency, the difference
between the J-C model and the exact model becomes smaller. And the
result also reveals that the frequency for atom decay and recover is
only dependent on the detuning $\delta$.

\begin{figure}
  \includegraphics{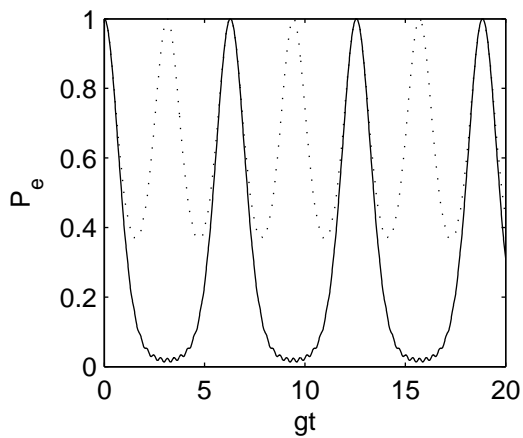}\\
  \caption{$P_{e}$ as a function of $gt$ for initially excited atom with $\omega_{0}=10g$. solid line for $\delta=0.2\omega_{0}$ and dotted line for $\delta=0.6\omega_{0}$.}\label{F4}
\end{figure}

\par (C) Here, the decay behavior of an atom in strong coupling regime is discussed.
For the detuning case, Fig.(\ref{F4}) shows that $P_{e}$ will
periodically decrease and recover, accompanying with small amplitude
rapid oscillation. As the detuning value increases, the oscillation
amplitude decreases, while the oscillation frequency increases.

\par Form Eq.(\ref{e7}) and Eq.(\ref{e71}), we could find that the contribution of energy-conserving process,
corresponding to rotating-wave terms $\sigma_{-}a^{\dagger}$ and
$\sigma_{+}a$ in Hamiltonian, varies with frequency of
$\omega_{0}-\omega$. And that of energy-non-conserving process,
corresponding to counter-rotating wave terms $\sigma_{+}a^{\dagger}$
and $\sigma_{-}a$ in Hamiltonian, varies with frequency of
$\omega_{0}+\omega$. The attenuation factor in Eq.(\ref{e10}) comes
from the destructive superposition of the different frequency
contribution associated with rotating-wave terms and
counter-rotating wave terms. With the increasing of coupling
strength, the contribution of virtual processes will increase and
result in the decay rate obviously modified.

\begin{figure}
  \includegraphics{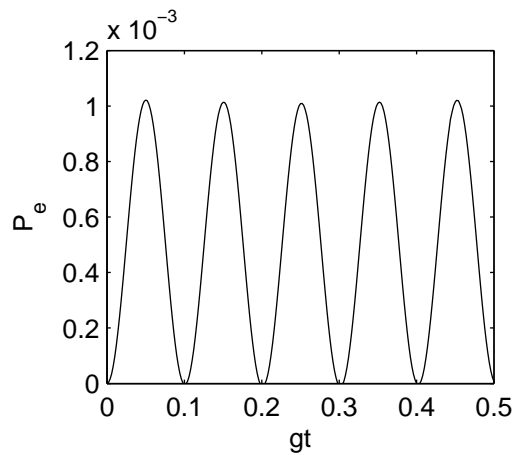}\\
  \caption{$P_{e}$ as a function of $gt$ for initially unexcited atom with $\omega_{0}=20g$ and $\delta=0.5\omega_{0}$.}\label{F5}
\end{figure}

\par Then, we focus on the case for initially unexcited atom. For J-C model, the
probability $P_{e}$ is always equal to zero, which indicates no Rabi
oscillation for an atom. However, the results for exact model is in
opposition to that.

\par For detuning case, Fig.(\ref{F5}) reals that probability $P_{e}$ periodically increases and
decays. It indicates that there is Rabi oscillation for initially
unexcited atom coupled with vacuum single-mode field, which is
contrary to that of J-C model. From another point of view, this
result theoretically testifies the existence of vacuum energy
fluctuation. The numeric result also exhibits that, as the coupling
strength enhances, the Rabi oscillation amplitude will increases.

\par From above discussion, we find that $P_{e}$ will oscillate with damping
amplitude for resonant case and there is Rabi oscillation for
detuning case. That is because spontaneous emission is inhibited if
there is detuning between atomic frequency and cavity mode, and
enhanced if the cavity is resonant\cite{purc,kle}.

\par In summary, we have investigated vacuum Rabi oscillation
 of an atom coupled with single-mode cavity field exactly, and compared
 the results with that of J-C model.
 \par Firstly, the results show that, for resonant case, there is
  damping Rabi oscillation for an atom, even in strong coupling regime,
 while there is Rabi oscillation for an tom in J-C model. Secondly, for small
 detuning and weak coupling case, the probability for the atom in excited
 state will oscillate against time with different frequency and amplitude
 from that of J-C model. Thirdly, the results also reveal that the
Rabi oscillation  frequency is only dependent on the detuning value,
while that is dependent on both the coupling strength and the
detuning value for J-C model. Fourthly, there is Rabi oscillation
for initially unexcited atom coupled with vacuum single-mode field,
which is contrary to that of J-C model.

\par On the whole, it exhibits that there is a significant effect on the dynamic behavior of the atom
 due to the counter-rotating wave interaction , even under the condition in which the RWA is considered
 to be justified. The vacuum energy fluctuation and counter-rotating
 wave interaction could invoke many different behaviors from that of
 J-C model.

\section*{Acknowledgments} This work was supported by the State
Key Program for Basic Research of China under Grant No.2007CB307001
and by the Science and Technology of Guangdong Province, China(Grant
No.2007B010400066).


\begin{thebibliography}{99}
\bibitem{Lou}W. H. Louisell, Quantum Statistical Properties of Radiation, (John Wiley \& Sons, New
York,1973).
\bibitem{pur}R.R. Puri, Mathematical Methods of Quantum Optics, Springer Series in Optical Sciences vol.79,(Springer-Verlag, Berlin, 2001).
\bibitem{kli}A. B. Klimov, I.Sainz and S.M.Chumakov, Phys. Rev. A 68,063811(2003).
\bibitem{rai}J. M. Raimond, M. Brune and S. Haroche, Rev. Mod. Phys.73,565(2001).
\bibitem{Ger}C. Gerry and P. Knight, Introductory Quantum Optics(Cambridge: Cambridge Univ.
Press, 2005 ).
\bibitem{mei}D. Meiser and P. Meystre, Phys. Rev. A 74,
065801(2006).
\bibitem{iri}E. K. Irish and K. Schwab, Phys.Rev. B 68, 155311(2003).

\bibitem{Ish}A. Ishizaki and Y. Tanimura, Chem. Phys. 347,185(2008).
\bibitem{zhang}H. X. Lu, J. Yang, Y. D. Zhang, and Z. B. Chen, Phys. Rev. A 67, 024101(2003).
\bibitem{purc}E. M. Purcell, Phys. Rev. 69, 681(1946).
\bibitem{kle}D. Kleppner, Phys. Rev. Lett. 47, 233(1981).
\end{thebibliography}
\end{document}